\documentclass{article}

\usepackage{arxiv}

\usepackage{xcolor}
\usepackage{subfigure}
\usepackage{siunitx}
\usepackage{booktabs}
\usepackage{amsmath,amsfonts,amssymb}
\usepackage{bm}
\usepackage{placeins}

\usepackage[utf8]{inputenc} 
\usepackage[T1]{fontenc}    
\usepackage[hidelinks]{hyperref}       
\usepackage{url}            
\usepackage{booktabs}       
\usepackage{nicefrac}       
\usepackage{microtype}      
\usepackage{lipsum}		
\usepackage{graphicx}
\usepackage{doi}

\let\vec\bm

\newcommand{\fTF}[3]{\ensuremath{\mathcal{F}_{#2 \mapsto #3}\!\left\{ #1 \right\}\!\left[#3\right]}}

\title{Machine-learning applied to classify flow-induced sound parameters from simulated human voice}
\renewcommand{\shorttitle}{ML to classify flow-induced sound}

\author{ \href{https://orcid.org/0000-0002-8652-5048}{\includegraphics[scale=0.06]{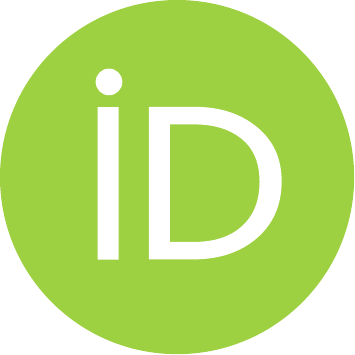}\hspace{1mm}Florian Kraxberger} \\
	Institute of Fundamentals and Theory in Electrical Engineering (IGTE)\\
	Graz University of Technology\\
	8010 Graz, Austria \\
	\texttt{kraxberger@tugraz.at} \\
	\And
	\href{https://orcid.org/0000-0002-8780-4073}{\includegraphics[scale=0.06]{orcid.pdf}\hspace{1mm}Andreas Wurzinger} \\
    Institute of Fundamentals and Theory in Electrical Engineering (IGTE)\\
	Graz University of Technology\\
	8010 Graz, Austria \\
	\texttt{andreas.wurzinger@tugraz.at} \\
	\And
	\href{https://orcid.org/0000-0002-2148-6703}{\includegraphics[scale=0.06]{orcid.pdf}\hspace{1mm}Stefan Schoder} \\
    Institute of Fundamentals and Theory in Electrical Engineering (IGTE)\\
	Graz University of Technology\\
	8010 Graz, Austria \\
	\texttt{stefan.schoder@tugraz.at} \\
}

\begin{document}
\maketitle

\begin{abstract}
Disorders of voice production have severe effects on the quality of life of the affected individuals. A simulation approach is used to investigate the cause-effect chain in voice production showing typical characteristics of voice such as sub-glottal pressure and of functional voice disorders as glottal closure insufficiency and left-right asymmetry. Therewith, 24 different voice configurations are simulated in a parameter study using a previously published hybrid aeroacoustic simulation model.
Based on these 24 simulation configurations, selected acoustic parameters (HNR, CPP, ...) at simulation evaluation points are correlated with these simulation configuration details to derive characteristic insight in the flow-induced sound generation of human phonation based on simulation results.
Recently, several institutions studied experimental data, of flow and acoustic properties and correlated it with healthy and disordered voice signals. Upon this, the study is a next step towards a detailed dataset definition, the dataset is small, but the definition of relevant characteristics are precise based on the existing simulation methodology of simVoice.
The small datasets are studied by correlation analysis, and a Support Vector Machine classifier with RBF kernel is used to classify the representations. With the use of Linear Discriminant Analysis the dimensions of the individual studies are visualized. This allows to draw correlations and determine the most important features evaluated from the acoustic signals in front of the mouth. 
The GC type can be best discriminated based on CPP and boxplot visualizations. Furthermore and using the LDA-dimensionality-reduced feature space, one can best classify subglottal pressure with 91.7\% accuracy, independent of healthy or disordered voice simulation parameters.
  
\end{abstract}



\section{Introduction}


Human phonation is the result of complex aeroacoustic phenomena such as periodical interruptions of the glottal airstream resulting in a pulsatile jet flow in the larynx, which are the main source of the human voice \cite{Titze2000}. Using numerical models enable to analyze the cause-effect chains in the complex human phonation system. Thereby, for the driven vocal folds simulation model \cite{Schoder2020} parameters constitute the 'cause' (i.e. regular and irregular oscillation configurations, sub-glottal pressure, ...), and the simulation results the 'effect' (i.e. flow field, acoustic pressure signals, ... ). According to \cite{Titze2000}, the efficiency of human phonation is mainly influenced by the vocal fold closure characteristics and the vocal fold motion properties. Thus, a high phonation efficiency is present, when the vocal folds close the glottis completely in each oscillation cycle, and when the vocal folds oscillate symmetrically as well as periodically. The phonation quality is reduced, when the glottis does not close completely or oscillates asymmetrically, resulting in a breathy or hoarse voice quality due to increased broadband components in the voice signal \cite{Park2008,Yamauchi2014,Hoffman2012}. These irregularities can occur in visually healthy larynges \cite{Rammage1992,Inwald2011,Patel2012} as well as pathologic larynges \cite{Schneider2003} and their occurrence correlates with increasing patient age \cite{Vaca2017,Soedersten1995}.

Recently machine learning-based approaches improved the insights in voice research \cite{moccia2018learning,cordeiro2017hierarchical,laves2019dataset} and to separate healthy from disordered voices \cite{callan1999self,awan2005acoustic,voigt2010classification}. 
Using a self-organizing map based on acoustic parameters, \cite{callan1999self} differentiated normal from disordered voices with 0.7628 accuracy. In \cite{awan2005acoustic} and for separation of normal, breathy, rough and hoarse voices using acoustic parameters 0.75 accuracy was achieved. Higher accuracy of 0.81, was achieved in \cite{voigt2010classification} using PVG based parameters. Also, other studies showed good prediction capabilities by using acoustic measures \cite{panek2015acoustic,umapathy2018automated}. Somewhat the pure differentiation between healthy and some features associated with functional dysphonia (for instance incomplete glottis closure) is not precise since they also occur in healthy subjects \cite{sama2001clinical} and some parameters are redundant \cite{schlegel2019dependencies,schlegel2019influence}. 
Regarding the broad range of possible impacts on the voice, experimental study data were enhanced with machine learning techniques to get a more profound understanding about the redundancy of which parameters best characterize functional dysphonia or other voice disorders \cite{schlegel2020machine}. 


Differently to experimental approaches (also using machine learning) and since disorders of voice production have severe effects on the quality of life, a simulation model \textit{simVoice} \cite{Falk2021} was developed to investigate the cause-effect chain in voice production with fully resolved data. The \textit{simVoice} model depicts typical characteristics of voice such as sub-glottal pressure and of functional voice disorders as glottal closure insufficiency and left-right asymmetry. In total up to 80 different voice configurations were simulated. This type of simulation results in high-resolution data that can be used for machine learning techniques as input. The motivation of this paper is to demonstrate the usability of and to perform systematic investigation on the voice signal (similar to \cite{umapathy2018automated}). The following simulation input parameters were used: (i) four different levels of glottis closure, (ii) symmetric and asymmetric vocal fold motion of the same frequency, and (iii) three different subglottal pressures, to analyze the relation between voice signal measures such as HNR, CPP and more. Also, we have to note that this is a first study on using simulation results for machine learning in the human voice community. Furthermore, the result interpretation is limited due to the small dataset. Nevertheless, interesting conclusions can be drawn from this high-quality, high-resolution simulation dataset.




\section{Methods}
\label{sec:02-methods}

In previous works, a validated hybrid aeroacoustic model of human phonation has been established \cite{Schoder2020,Schoder2021,Sadeghi2019b, Sadeghi2019}, which is called \textit{simVoice}. To reduce the computational cost of a fluid-structure-acoustic interaction model \cite{Sadeghi2019}, a prescribed vocal fold motion model is used \cite{Sadeghi2019a}.
This model consists of a finite-volume incompressible fluid dynamics simulation, and a finite-element acoustic simulation, as documented in \cite{Maurerlehner2021}. In \cite{Falk2021}, the model has been used to model different regular and irregular phonation characteristics, such as various degrees of insufficient glottal closure, symmetric or asymmetric vocal fold motion, and different subglottal pressures. This section is a brief introduction to the numerical model \textit{simVoice} and the acoustic features of the simulated voice signals are defined. The aim is to discriminate against different healthy and functional-based voice disorders solely based on the evaluated acoustic features (e.g. \cite{umapathy2018automated}). This approach originates in voice and music research and usually uses a large number (often $N_\text{features} > 20$) of audio features \cite{Alias2016,Eyben2016,Geiger2013,Mierswa2005}. Regarding the small dataset and the often redundant features \cite{schlegel2019dependencies}, only a few features were selected for the present study are: (i) sound pressure level (SPL), (ii) Harmonics-to-Noise Ratio (HNR), (iii) Cepstral Peak Prominence (CPP), (iv) spectral slope, (v) Hammarberg Index (HBI), and (vi) Alpha Ratio (see Sec.~\ref{sec:feature-definition}). 

\subsection{Hybrid Aeroacoustic Simulation Setup for Human Phonation (\textit{simVoice})}

The hybrid aeroacoustic simulation setup used to model human phonation consists of four components: (i) an incompressible computational fluid dynamics (CFD) simulation, (ii) conservative interpolation from the dense grid of the CFD to the coarser grid of the computational acoustics (CA) simulation, (iii) computation of the aeroacoustic source term and (iv) acoustics simulation based on the Perturbed Convective Wave Equation (PCWE).

\begin{figure}[htbp]
\centering
\includegraphics[width=0.5\linewidth]{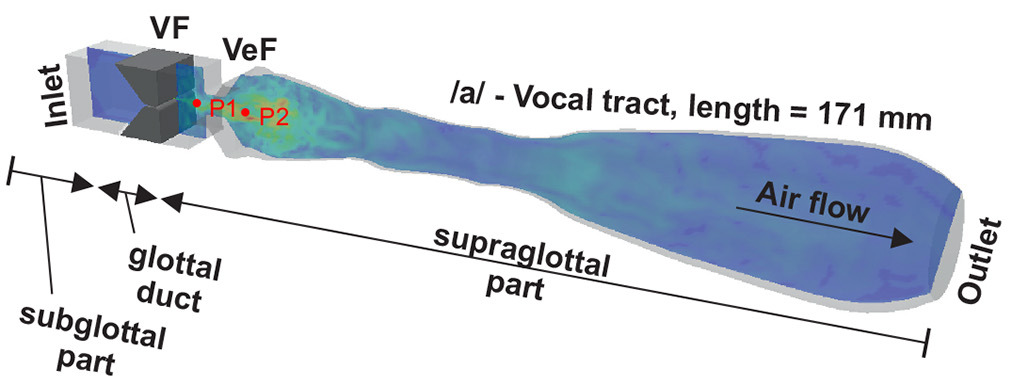}
\caption{(color online) Computational domain of the CFD. The subglottal pressure is applied as a constant inlet pressure boundary condition \cite[Fig.~2(A)]{Falk2021}.}
\label{fig:CFD-domain}
\end{figure}

The incompressible CFD simulation is documented in \cite{Sadeghi2019b,Sadeghi2019a,Sadeghi2019} and has been performed with the software \textit{StarCCM+}. The result thereof is the incompressible pressure $p^{\mathrm{ic}}$, as described in \cite{Falk2021}, and \cite{Schoder2021} provides an in-depth analysis of the aeroacoustic source terms. The aeroacoustic source terms are furthermore filtered with the algorithm proposed in \cite{Schoder2021a}. Figure~\ref{fig:CFD-domain} shows the computational domain of the CFD simulation.

The PCWE, as defined in \cite{Kaltenbacher2015,Schoder2020}, is a wave equation regarding the acoustic scalar potential $\psi^\mathrm{a}$. 
From the acoustic scalar potential $\psi^{\mathrm{a}}$, the acoustic pressure $p^\mathrm{a}$ is evaluated. 

\begin{figure}[htbp]
	\centering
	\subfigure[Slice of the CA domain]{\includegraphics[height=3.5cm,keepaspectratio]{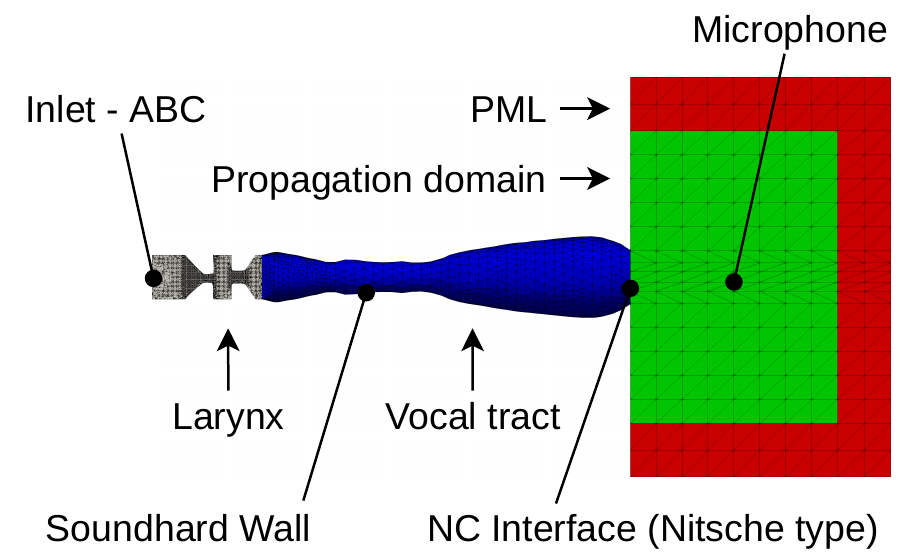}}
	\subfigure[CA domain]{\includegraphics[height=3.5cm,keepaspectratio]{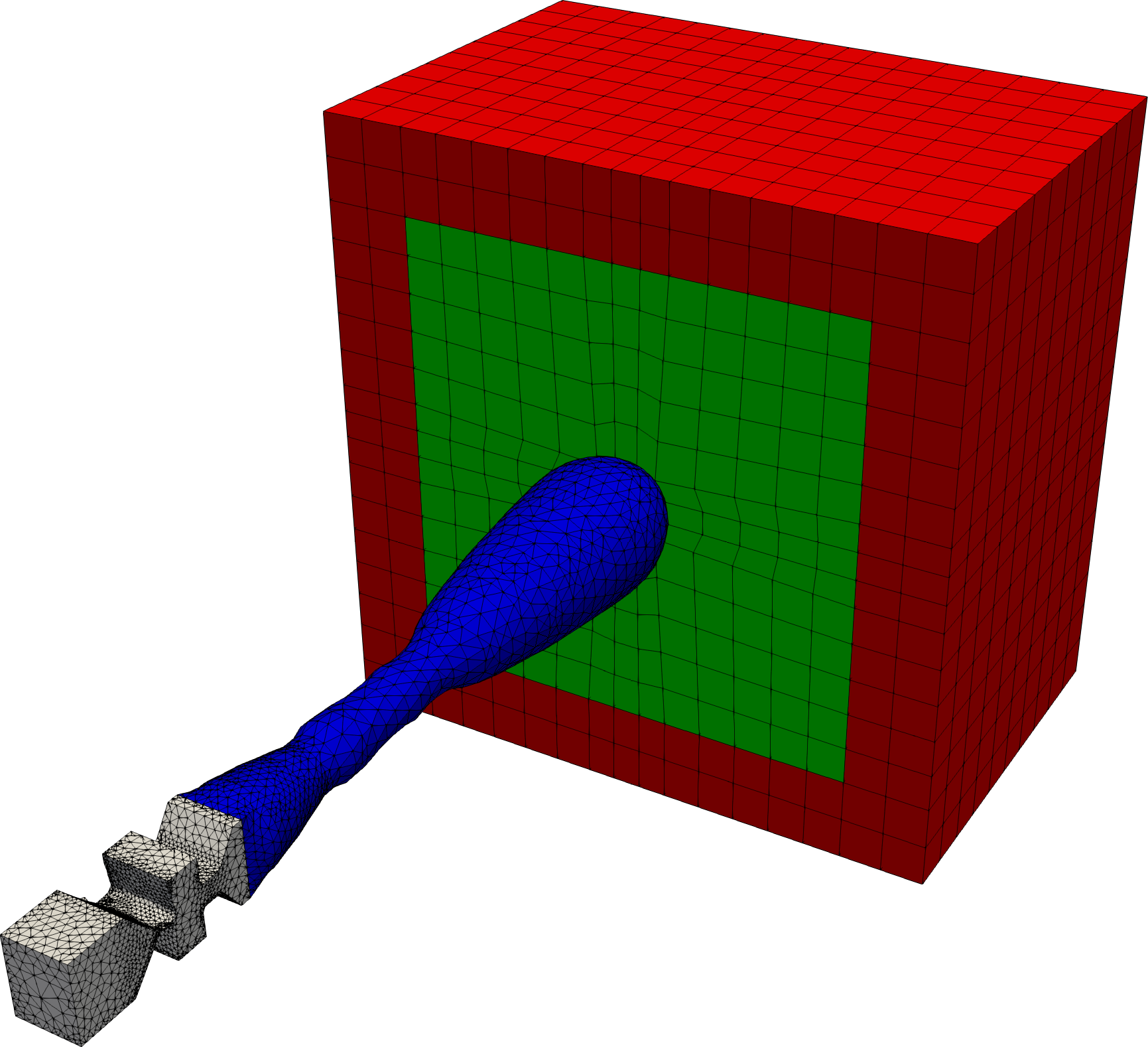}}
	\caption{(color online) Computational domain of the CA simulations.
    }
	\label{fig:02-CA}
\end{figure}

It has been shown in \cite{Schoder2021}, that the convective term can be neglected in the substantial derivative, without a significant deviation in the spectrum of the resulting acoustic pressure. The PCWE is solved using the finite-element solver \textit{openCFS} \cite{schoder2022opencfs}. An in-depth description of the CA simulations is available in \cite{Maurerlehner2021, Schoder2020}. Fig.~\ref{fig:02-CA} shows the domain of the CA simulations consisting of the larynx, vocal tract, propagation and perfectly matched layer (PML) regions. The CA simulation domain presents an extension of the CFD simulation domain shown in Fig.~\ref{fig:CFD-domain} with a propagation and PML region. For the inlet surface, an absorbing boundary condition (ABC) was used in the CA simulations. The vocal tract and propagation regions are connected with a non-conforming (NC) interface, as described in \cite{Maurerlehner2021}. The PML region ensures free-field wave propagation for the propagation domain \cite{Kaltenbacher2013}. The walls of the vocal tract are modeled sound-hard.

\subsection{Phonation configurations (labels)}
\label{sec:02-voice-disorders}
Voice disorders are phenomena containing complex interactions between different parts of the voice production process, such as the subglottal pressure coming from the bronchial tubes, the vocal folds (oscillating at a frequency of \SI{148}{\hertz} representing male voice), the false vocal folds, the vocal tract (/a/ \cite{Probst2019}). For the present study, these interacting elements and regular (subglottal pressure) and irregular characteristics (asymmetry and insufficiency) thereof are abstracted in terms of certain functions of the voice production process (see \cite{Falk2021}). Thus, periodic symmetric and asymmetric oscillation characteristics, different grades of glottis insufficiency, in combination with varying subglottal pressure have been simulated as described in \cite{Falk2021,Falk2021a}. The simulation results of the acoustic pressure signal in front of the vocal tract (the mouth), are used for the later computations of the acoustic features. 

\paragraph{Subglottal Pressure}
The presence of a pressure difference between the subglottal and supraglottal regions drives the glottal airstream. The inability to build up such a pressure can be one reason for a dysphonic voice, e.g. caused by muscle tension disorders, as reported in \cite{Belsky2021,Garaycochea2019}.
The simulation cases include the subglottal pressures as follows
\begin{equation}
    P_\text{inlet}\in\left\{ 385, 775, 1500 \right\}\SI{}{\pascal}.
\end{equation}
Where, $\SI{775}{\pascal}$ represents normal subglottal pressure and the other low or high, respectively.

\paragraph{Glottal Closure Type}
The vocal folds can have varying degrees of glottal insufficiency, which occur when the vocal folds do not close completely in each cycle. According to \cite{Kniesburges2020, Patel2014}, this irregularity occurs for many voice pathologies. As a consequence, concerned patients report, that a higher effort is necessary for phonation \cite{Zhang2019a}.
The glottal closure (GC) type determines the amount of glottal insufficiency. $o_\mathrm{initial}$ is defined as the percentage of the vocal folds lengths which does not close. The simulation model employs initial glottal openings of
\begin{equation}
    o_\mathrm{initial}\in \left\{ 0 \text{(GC1)}, 40, 70, 100 \text{(GC4)} \right\}\SI{}{\percent},
\end{equation}
which are called GC1 (complete glottal closure), GC2 (60\% of the vocal folds lengths close), GC3 (30\% of the vocal folds lengths close), and GC4 (no contact of vocal folds during oscillations), respectively \cite{Falk2021}. The different GC types are based on high-speed videos from experiments with excised larynges and are motivated by clinical observations \cite{Lodermeyer2015, Birk2017, Sadeghi2019}. A detailed description as well as visualizations of the different symmetry cases are available in Fig.~5.5~\cite{Maurerlehner2020}.

\begin{figure}[htb]
\centering
\includegraphics[width=0.45\linewidth]{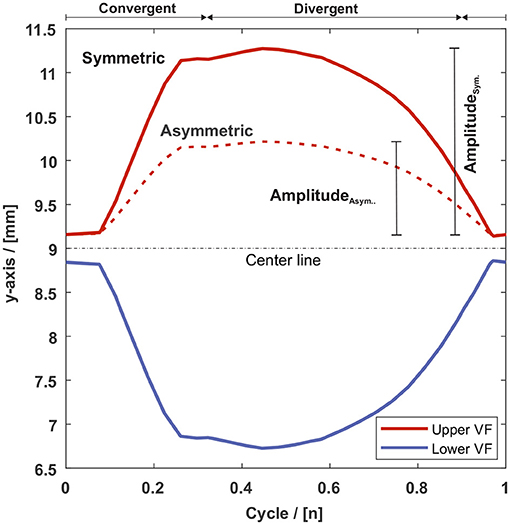}
\caption{(color online) Exemplary vocal fold motion of GC1 for the symmetric and asymmetric case along the y-axis (medial-lateral direction) for the center point on the medial plane of the VF surface \cite[Fig.~4]{Falk2021}.}
\label{fig:sym-asym-motion}
\end{figure}

\paragraph{Vocal Fold Motion Symmetry}
In regular phonation, the left and right vocal fold oscillate symmetrically. However, in the case of laryngeal unilateral paresis or unbalanced muscle tension, the vocal folds oscillate  asymmetrically \cite{Woo2016Analysis, Azar2021Perceptual}.
Therefore, the proposed simulation model features two cases of vocal fold motion symmetry. In the symmetric case, both vocal folds have the same amplitude in their motion. In the asymmetric case, one vocal fold oscillates with \SI{50}{\percent} of the amplitude of the other vocal fold. Both symmetry cases are depicted in Fig.~\ref{fig:sym-asym-motion}.

In Tab.~\ref{tab:02-simulation-parameters}, the simulation parameters are summarized. A total of 24 different functional-based voice disorders have been simulated.

\begin{table}[htbp]
\centering
\caption{Simulation parameters of functional-based voice disorders as defined in \cite{Falk2021}.}
\begin{tabular}{@{}lrl@{}}
\toprule
\textbf{Parameter} & \# & \textbf{Values} \\
    \midrule
    Subglottal Pressure & 3 & $P_\text{inlet}\in\left\{ 385, 775, 1500 \right\}\SI{}{\pascal}$ \\
    Glottal Closure Type & 4& $o_\mathrm{initial}\in \left\{ 0, 40, 70, 100 \right\}\SI{}{\percent}$  \\
    Vocal Fold Motion Symmetry & 2& asymmetric or symmetric  \\
    \midrule
    & 24 & configurations total  \\
    \bottomrule
\end{tabular}
\label{tab:02-simulation-parameters}
\end{table}

\subsection{Acoustic Feature Definitions}
\label{sec:feature-definition}

The acoustic pressure signal results of the \textit{simVoice} model are recorded at an evaluation point (so-called 'microphone'), as depicted in Fig.~\ref{fig:02-CA}~(a). The recorded signal is time discrete acoustic pressure $p[n]$ with $n$ being the time index \cite{Oppenheim2014} sampled at $T_s = 1/f_s = \SI{13.6}{\micro\second}$ \cite{Schoder2020, Falk2021}. With those audio signals (one resolved signal for each of the 24 simulation configurations), we compute certain \textit{acoustic features} for the analysis. Thus, we aim to analyze the flow-induced sound generation of health and of functional-based voice disorders by using the evaluated features only.

\paragraph{Sound Pressure Level}
The SPL $L_p$ of an acoustic pressure signal $p[n]$ is computed as follows,
\begin{equation}
\begin{split}
    L_p &= 20 \cdot \log_{10}{ \frac{\tilde{p}}{p^\mathrm{a}_\mathrm{ref}}} \qquad \text{with}  \qquad p^\mathrm{a}_\mathrm{ref} = \SI{20}{\micro\pascal} \\
    \tilde{p} &= \sqrt{\frac{1}{N} \sum_{n=1}^{N} \left( p[n] \right)^2}.
\end{split}
\end{equation}
Thereby, $N$ is the number of time steps of $p[n]$. Consequently, $\tilde{p}$ denotes the root-mean-square of the acoustic pressure signal. The subglottal pressure is expected to have a significant influence on the sound pressure level (SPL) magnitude of the acoustic signal. Since a higher subglottal pressure (driving force of the phonation) will certainly exhibit a stronger voice at the same vocal efficiency.

\paragraph{Harmonics-to-Noise Ratio}
Harmonics-to-Noise ratio (HNR) provides a measure for the degree of hoarseness \cite{Yumoto1982}, and has since been used, among others, to identify voice disorders \cite{Lee2018} and voice quality \cite{Kreiman2014,Ferrand2002}.
Thus, in the present study HNR is used with the aim of contributing to the identification and analysis of glottis closure type, the asymmetry and their combinations defined in Sec.~\ref{sec:02-voice-disorders}.
The idea is to use the HNR to be a measure of an increasing amount of turbulent structures and decreasing tonal contribution during the whole oscillation cycles for only partly closing vocal folds.
HNR is defined as the energy ratio between the harmonic signal component to the noise-like signal component in \SI{}{\decibel}. The splitting of harmonic and noise-like components is performed via the auto-correlation function (ACF), as described in \cite[pp.~77--78]{Eyben2016}. The ACF $r_{p}(\tau)$ of a pressure signal $p[n]$ is calculated with
\begin{equation}
    r_{p}(\tau) = \sum_{n=0}^{N-1} p[n]p[n+\tau],
\end{equation}
where $\tau$ is the independent lag variable. Therewith, the HNR can be computed with \cite[Eq.~(2.234)]{Eyben2016}, such that
\begin{equation}
\mathrm{HNR}_\mathrm{dB,Eyben} = 10  \log_{10}{\frac{ \max\left\{ r_{p}(\tau) \right\} }{ r_{p}(0) - \max\left\{ r_{p}(\tau) \right\}}},
\end{equation}
under the assumption that (i) the noise is additive, (ii) it is uncorrelated with the signal and (iii) the noise is uncorrelated to itself (i.e. it is white noise). To be robust against deviations from the assumptions in the simulated signal, a margin of $k_\mathrm{marg}=300$ lag bins is introduced, which corresponds to a lag of $\tau_\mathrm{marg} = k_\mathrm{marg} / f_s = \SI{4.08}{\milli\second}$. Therewith, the HNR $h$ is calculated as follows:
\begin{equation}
\mathrm{HNR}_\mathrm{dB} = h = 10  \log_{10}{\frac{ \max_{\tau_\mathrm{marg}}^{\tau_\mathrm{max}}\left\{ r_{p}(\tau) \right\} }{ r_{p}(0) - \max_{\tau_\mathrm{marg}}^{\tau_\mathrm{max}}\left\{ r_{p}(\tau) \right\} }},
\end{equation}
where $\tau_\mathrm{max}$ is the maximum possible lag time (i.e. the time corresponding to the highest lag index).

\paragraph{Cepstral Peak Prominence}
The harmonic structures occurring in the voice spectrum map to peaks in the cepstral domain, as introduced in \cite{Bogert1963}. Healthy voice shows, that the harmonic parts dominate the spectrum, and thus it is expected that the corresponding peak in the cepstral domain is prominent compared to non-harmonic parts of the cepstrum. Hence, Cepstral Peak Prominence (CPP) measures the prominence of the peak in the cepstral domain and therefore the prominence of harmonic structures in the spectrum compared to non-harmonic parts. Therewith, we aim to discriminate simulation parameters which affect the harmonic structure of the voice signal, e.g. by an incomplete glottis closure.
The CPP is computed as follows \cite{Hillenbrand1994, Hillenbrand1996}: Given a pressure signal $p[n]$, a power spectral density (PSD) estimate $\tilde{p}[k]$ is computed using Welch's method \cite{Welch1967} such that
\begin{equation}
\begin{split}
\tilde{p}[k] &= \sqrt{\fTF{p[n]}{n}{k}} \\
\tilde{p}_\mathrm{dB}[k] &= 20 \cdot \log_{10}{\frac{\tilde{p}[k]}{p_0}} \qquad \text{with} \qquad p_0=\SI{20}{\micro\pascal},
\end{split}
\end{equation}
where $\fTF{p[n]}{n}{k}$ denotes the PSD based on the $N$-point discrete Fourier transform (DFT) \cite{Oppenheim2014}. In accordance to \cite{Bogert1963}, the cepstrum $c_{\tilde{p}}[k]$ is computed such that
\begin{equation}
c_{\tilde{p},\mathrm{dB}}[q] = 10\cdot \log_{10}{\left| \fTF{\tilde{p}_\mathrm{dB}[k]}{k}{q}  \right|^2}.
\end{equation}
The independent variable of the cepstrum is called \textit{quefrency} and has the dimension of \textit{time} \cite{Bogert1963}. The CPP $c$ is evaluated as follows \cite{Hillenbrand1994, Hillenbrand1996}: For quefrencies larger than \SI{1}{\milli\second}, a linear regression line is fitted to the cepstrum. The CPP is the difference between the peak amplitude of the cepstral maximum and the linear regression line evaluated at the quefrency of the cepstral maximum. Thus, it represents how far the cepstral peak emerges from the cepstral "background noise". This can be interpreted as the prominence of harmonics in the signal compared to non-harmonic signal portions. In \cite{Birk2017,Rosenthal2014}, it was used together with other features as a measure for vocal effort and to disambiguate different voice disorders. In \cite{Maryn2010a,Maryn2010,Maryn2009}, CPP and HNR are used to compute the so-called \textit{Acoustic Voice Quality Index} together with shimmer and spectral slope and tilt.
Furthermore, CPP and cepstrum-based features are used as fundamental measures in speech sciences \cite{Behrman2021,Ellis2010,Gold2011}.


\paragraph{Spectral Slope}
The spectral slope is a measure of energy distribution across the spectrum and is used in \cite{Kreiman2012, Chen2011} to distinguish voice qualities. If the glottal airstream is interrupted off completely in each vocal fold oscillation cycle, we expect effects on the energy distribution in the resulting acoustic signal. Thus, this feature is used to determine if high frequencies (which we expect to increase due to incomplete glottis closures) dominate lower frequencies, such that the general spectral trend represented by the spectral slope is influenced.
In \cite{Eyben2016}, spectral slope\footnote{As shown in \cite{Kakouros2018}, the terms "spectral slope" and "spectral tilt" are used interchangeably in the literature, thus, for spectral slope, the definition of \cite[pp.~35]{Eyben2016} is used.} is defined as the result of a linear regression operation on the whole spectrum $\tilde{p}[k]$. Thereby, a line $\hat{y} = ax + b$ is fitted to the spectrum $\tilde{p}[k] =\sqrt{ \fTF{p[n]}{n}{k}}$, such that the mean-squared error between the line and the original spectrum is minimized (\textit{minimum mean-squared error} condition). The parameter of interest is $a$, which is the \textit{slope} of the linear regression result. A formula for $a$ is provided by \cite{Eyben2016}, such that
\begin{equation}
a = \frac{N \Sigma_{xy} - \Sigma_x \Sigma_y}{N \Sigma_{x^2}-\Sigma_x^2},
\end{equation}
where $N$ is the number of DFT frequency bins and $\Sigma_{xy}$, $\Sigma_x$, $\Sigma_y$ and $\Sigma_{x^2}$ are defined as follows
\begin{equation}
\begin{split}
\Sigma_x = \sum_{i=0}^{N-1} f(k_i),  &\qquad \Sigma_y =\sum_{i=0}^{N-1} p[i],  \\
\Sigma_{x^2}= \sum_{i=0}^{N-1} f(k_i)^2,  &\qquad \Sigma_{xy}=\sum_{i=0}^{N-1} f(k_i) p[i],
\end{split}
\end{equation}
where $f(k)$ is the frequency in \SI{}{\hertz} associated with the frequency bin $k$. By exploiting the equidistant frequency scale used in the DFT, simplifications arise \cite{Eyben2016}, such that
\begin{equation}
\Sigma_x = \frac{1}{2} N(N-1) \qquad \text{and} \qquad \Sigma_{x^2} = \frac{1}{6} N(N-1)(2N-1).
\end{equation}

\paragraph{Hammarberg Index and Alpha Ratio}
Hammarberg Index (HBI) and Alpha Ratio are measures of the energy distribution across the spectrum. Similar to spectral slope, also these features are introduced aiming to discriminate simulation parameters which influence the glottis closure, i.e. GC type and glottis symmetry.
HBI $\eta$ as defined in \cite{Hammarberg1980} is the ratio of the most prominent energy peak in the range of \SI{0}{\kilo\hertz}--\SI{2}{\kilo\hertz} to the most prominent energy peak in the region \SI{2}{\kilo\hertz}--\SI{5}{\kilo\hertz}. It is calculated according to \cite[p.~38]{Eyben2016}, such that
\begin{equation}
\eta = \frac{\max_{k=1}^{k_\mathrm{pivot}}\left\{\tilde{p}[k]\right\}}{\max_{k=k_\mathrm{pivot}+1}^{k_\mathrm{max}}\left\{\tilde{p}[k]\right\}},
\end{equation}
where the pivot frequency bin $k_\mathrm{pivot}$ is the highest spectral bin for which $f\leq \SI{2}{\kilo\hertz}$, and $k_\mathrm{max}$ is the highest spectral bin for which $f\leq \SI{5}{\kilo\hertz}$.

The alpha ratio is similar to the HBI described above, but instead of computing the ratio between energy peaks, the energy sum in the frequency bands is considered, as introduced in \cite{Patel2010}. Using the frequency bands \SI{50}{\hertz}--\SI{1}{\kilo\hertz} and \SI{1}{\kilo\hertz}--\SI{5}{\kilo\hertz}, the alpha ratio $\rho_\alpha$ can be computed according to \cite[pp.~38--39]{Eyben2016}, such that
\begin{equation}
\rho_\alpha = \frac{\sum\limits_{k=k_\mathrm{start}}^{k_\mathrm{pivot}}\tilde{p}[k]}{\sum\limits_{k=k_\mathrm{pivot}+1}^{k_\mathrm{max}}\tilde{p}[k]},
\end{equation}
where the $k_\mathrm{start}$ is the lowest spectral bin for which $f\geq \SI{50}{\hertz}$, $k_\mathrm{pivot}$ is the highest spectral bin for which $f\leq\SI{1}{\kilo\hertz}$, and $k_\mathrm{max}$ is the highest spectral bin for which $f\leq \SI{5}{\kilo\hertz}$.


\paragraph{Low-Pass Filtering}

The acoustic pressure signals from the simVoice model is filtered by low-pass filter. Before the computation of all features a filter with \SI{5}{\kilo\hertz} cut-off frequency is used. This is due to the upper frequency limit of the simulation setup \cite{Schoder2020, Schoder2021a}.
In addition, and before computing CPP, HNR and spectral slope, we use a filter of \SI{2}{\kilo\hertz} cut-off frequency targeting only the lowest two formants \cite{Probst2019}. The features based on both filtering strategies in total amount to 9 evaluated acoustic  features per configuration $i$.
The evaluated features for the $i$-th configuration ($i=1,\dots,24$) are denoted as a vector $\vec{F}^{(i)}$, such that
\begin{equation}
\vec{F}^{(i)} = \begin{bmatrix}
    L^{(i)}_{P,\SI{5}{\kilo\hertz}} & h^{(i)}_{\SI{5}{\kilo\hertz}} & h^{(i)}_{\SI{2}{\kilo\hertz}} & c^{(i)}_{\SI{5}{\kilo\hertz}} & c^{(i)}_{\SI{2}{\kilo\hertz}} & a^{(i)}_{\SI{5}{\kilo\hertz}} & a^{(i)}_{\SI{2}{\kilo\hertz}} & \eta^{(i)}_{\SI{5}{\kilo\hertz}} & \rho_{\alpha,\SI{5}{\kilo\hertz}}^{(i)}
    \end{bmatrix}^\mathrm{T},
\end{equation}
where $L_P^{(i)}$ is SPL, $h^{(i)}$ is HNR, $c^{(i)}$ is CPP, $a^{(i)}$ is spectral slope, $\eta^{(i)}$ is HBI, $\rho_\alpha^{(i)}$ is the alpha ratio, and the subscript indicates the low-pass filter's cut-off frequency.

\subsection{Data Labeling, Dimensionality Reduction and Classification}
\label{sec:02-LDA-SVM}
\begin{figure}[htbp]
\centering
\includegraphics[width=0.5\linewidth]{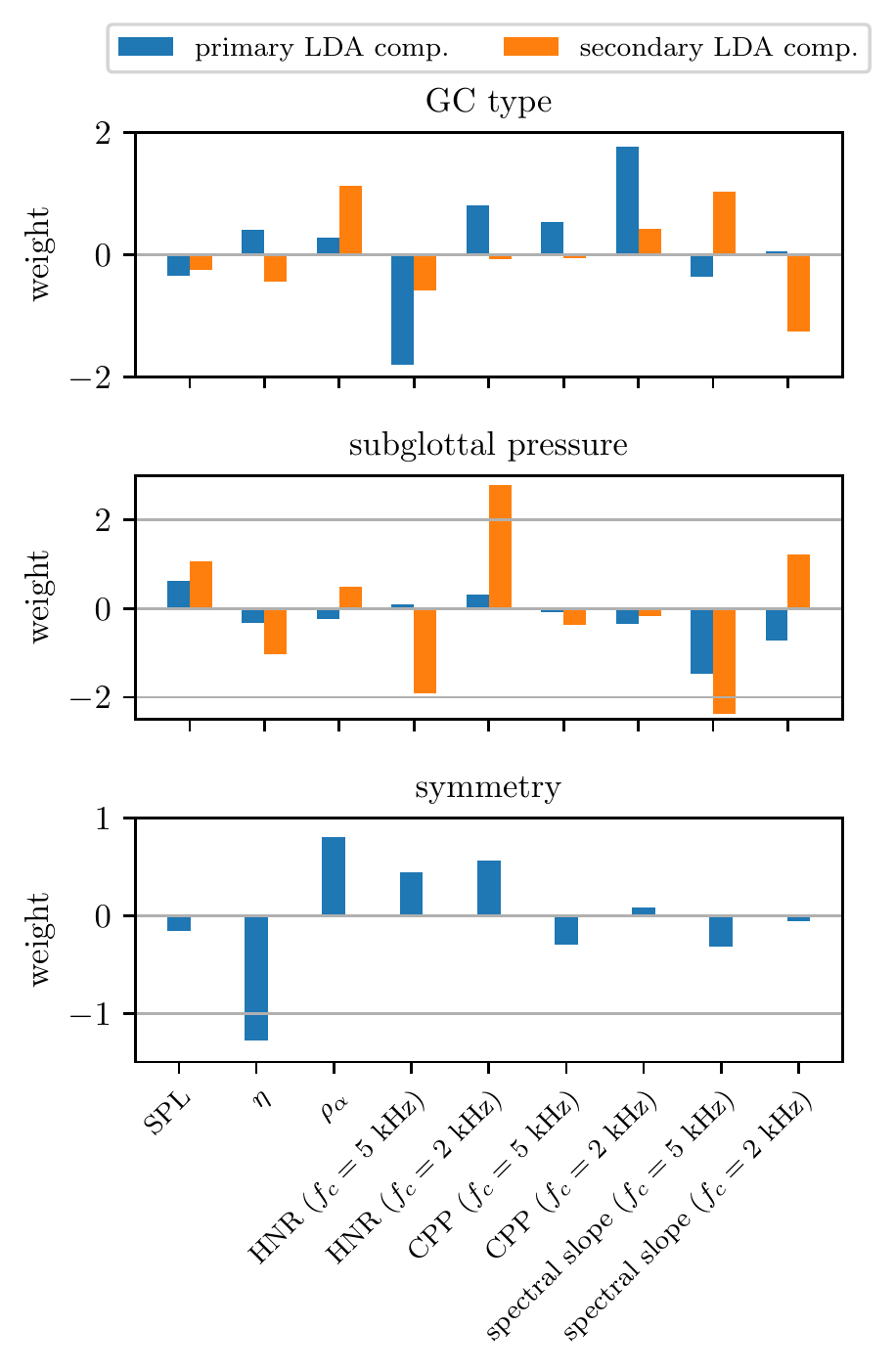}
\caption{Feature weights for dimension reduction.}
\label{fig:02-LDA-weights}
\end{figure}

We interpret different functional-based voice disorders as \textit{labels} in a machine learning sense. Thus, each of the 24 configurations defined in Sec.~\ref{sec:02-voice-disorders} is labeled with one value of each of the following label sets. For the $i$-th configuration, this yields labels from the following label sets:
\begin{equation}
\begin{split}
    \ell^{(i)}_{\mathrm{subglottPress}} &\in \left\{ 385, 775, 1500  \right\} \\
    \ell^{(i)}_{\mathrm{GC}} &\in \left\{ 1, 2, 3, 4 \right\} \\
    \ell^{(i)}_{\mathrm{sym}} &\in \left\{ 0, 1 \right\} .
\end{split}
\end{equation}
Thus, the label $\vec{L}^{(i)}$ of the $i$-th simulation configuration is formed as a vector
\begin{equation}
\begin{split}
    \vec{L}^{(i)} = \begin{bmatrix}
    \ell^{(i)}_{\mathrm{subglottPress}} & \ell^{(i)}_{\mathrm{GC}} & \ell^{(i)}_{\mathrm{sym}}\end{bmatrix}^\mathrm{T}
\end{split}
\end{equation}
For example, considering the configuration with \SI{385}{\pascal} subglottal pressure, GC1 and asymmetric VF motion, the label vector would be $\vec{L}^{(1)} = \begin{bmatrix}
    385 & 1 & 0
    \end{bmatrix}^\mathrm{T}$.
To form classes, different oscillation patterns and subglottal pressure are considered individually. Thereby, all evaluated feature vectors $\vec{F}^{(i)}$ are considered to be of a certain class, if \textit{one} element of the feature vector $\vec{L}^{(i)}$ is identical. The other elements of the feature vector are discarded. For example, the class GC1 contains all data samples $\vec{F}^{(i)}$, for which $\vec{L}^{(i)} = \begin{bmatrix}
    \ell^{(i)}_{\mathrm{subglottPress}} & 1 & \ell^{(i)}_{\mathrm{sym}}
    \end{bmatrix}^\mathrm{T}$,
regardless of the values of $\ell^{(i)}_{\mathrm{subglottPress}}$ and $\ell^{(i)}_{\mathrm{sym}}$.
The features, as defined in Sec.~\ref{sec:feature-definition}, span a 9-dimensional feature space.
For dimension reduction and visualization, a two-fold approach is used, consisting of (i) Linear Discriminant Analysis (LDA), see Sec.~4.1.4 of \cite{Bishop2006} and (ii) upon the results of the LDA a classification with Support Vector Machines (SVMs), see Sec.~7.1.3 of \cite{Bishop2006}. LDA is used to project the data points $\vec{F}^{(i)}$ from the original feature space into a two-dimensional feature space, with the goal to maximize the distance between different classes while minimizing the covariance within each class. The results of the LDA are transformed data points $\vec{F}^{(i)}_\mathrm{tf}$ for which $\mathrm{dim}\!\left(\vec{F}^{(i)}_\mathrm{tf}\right)=2$ for the LDA with respect to GC type (only using the first two major components for visualization) and subglottal pressure, and $\mathrm{dim}\!\left(\vec{F}^{(i)}_\mathrm{tf}\right)=1$ for the LDA with respect to symmetry property.
In Fig.~\ref{fig:02-LDA-weights}, the weights of the LDA transformation are depicted.

Thereafter, the labeled data is used to train an SVM classifier with radial basis function-kernels and a one-versus-rest classification. The SVM implementation in the Python library \textit{sklearn} has been used for this task \cite{Pedregosa2011}.
To quantify the generalization ability of the SVM classifier, $5$-fold stratified cross-validation was used \cite{Bishop2006}. The mean scores of the cross-validation data sets are shown in Tab.\ref{tab:03-acouPress-SVM-scores}. Thereafter, the whole data set was used to train the SVM classifier to classify mesh grid points to visualize the class boundaries. Based on the small number of data, the cross-validation procedure leads to stratified results but should be considered in the future with increasing amount of data.

\FloatBarrier
\section{Results}
\label{sec:03-results}

\begin{figure}[htb]
\centering
\includegraphics[width=0.7\linewidth]{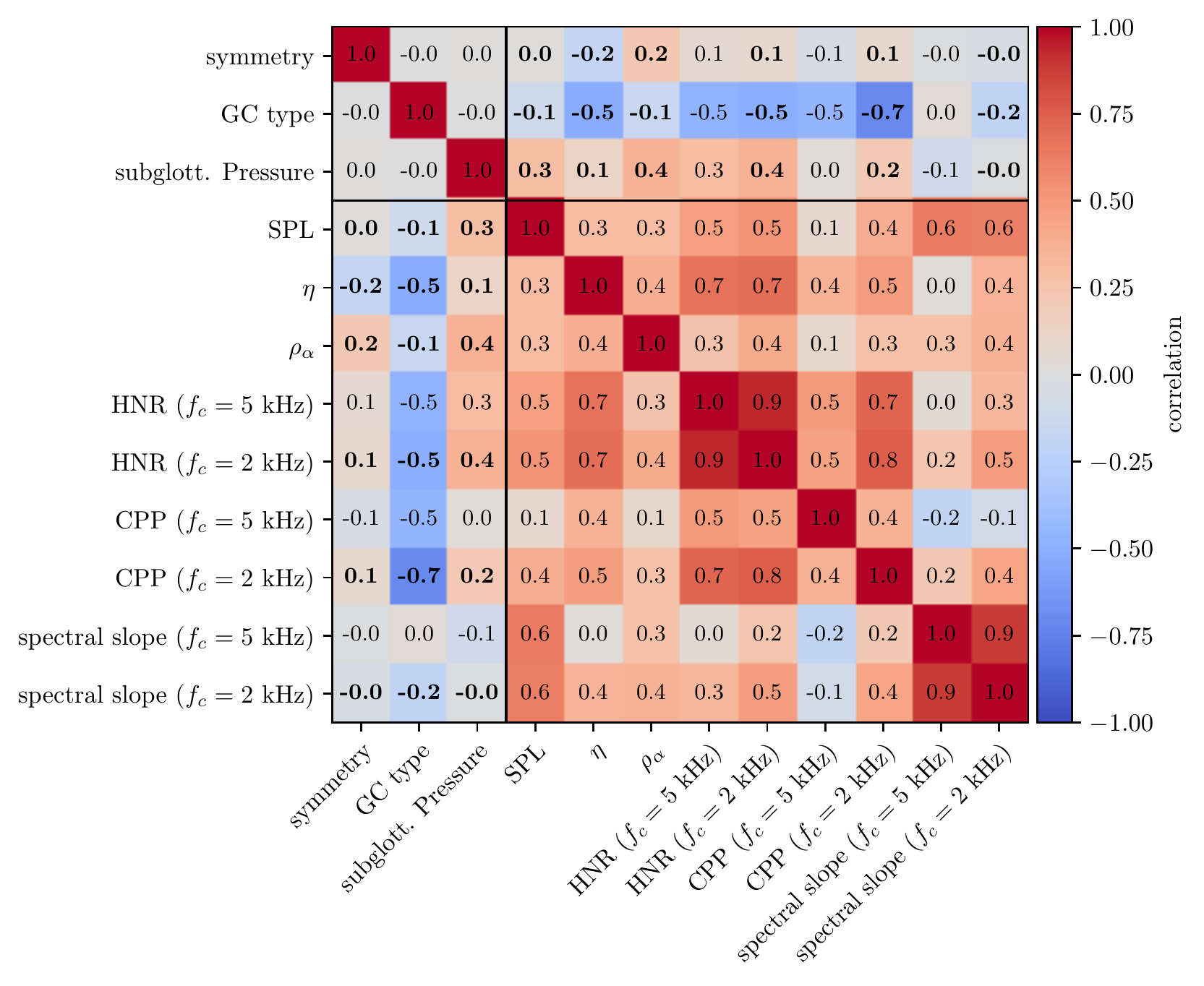}
\caption{
(color online) Correlation map between features and parameters for signals (acoustic pressure at the microphone position). The fields are colored according to the correlation value of the respective parameter combination (i.e. the number in each field). The numbers in bold highlight cause-effect-combinations which are further investigated using boxplot diagrams. The thick black lines are separating simulation parameters from evaluated features.}
\label{fig:03-vis-acouPress-periodic-corrMap}
\end{figure}

First, we study Pearson's correlation between the individual acoustic features and display it as a correlation map. In Fig.~\ref{fig:03-vis-acouPress-periodic-corrMap}, the correlation map is shown for configurations evaluated from acoustic pressure at the evaluation position as depicted in Fig.~\ref{fig:02-CA}~(a). 


From Fig.~\ref{fig:03-vis-acouPress-periodic-corrMap} it is visible, that for HNR, CPP and spectral slope, which have been evaluated following a \SI{2}{\kilo\hertz} LP-filter provide stronger correlations than those evaluated with a \SI{5}{\kilo\hertz} LP-filter.
Furthermore, it is visible that spectral slope does not correlate significantly with any of the modeled phonation characteristics, i.e. symmetry, GC type, and subglottal pressure. Based on the small dataset, the significance of the individual correlation based on the autocorrelation is not giving a full picture. Therefore, we continue studying the data with boxplots to get a more detailed understanding of a significant difference of the data points.

\FloatBarrier

\subsection{Boxplot Visualizations}

In the following, boxplot diagrams are shown, which visualize the distributions of the evaluated features. Thereby, manually selected 6 features ($L^{(i)}_{P,\SI{5}{\kilo\hertz}}, h^{(i)}_{\SI{2}{\kilo\hertz}}, c^{(i)}_{\SI{2}{\kilo\hertz}}, a^{(i)}_{\SI{2}{\kilo\hertz}}, \eta^{(i)}_{\SI{5}{\kilo\hertz}}, \rho_{\alpha,\SI{5}{\kilo\hertz}}^{(i)}$) are used because the correlation diagram in Fig.~\ref{fig:03-vis-acouPress-periodic-corrMap} already showed, which features provide better correlations with respect to the simulation parameters.  From the boxplot diagrams, we can learn, how individual features vary across different simulation parameters. They provide an insight on possible parameter groups which cannot be discriminated using individual features, but might be discriminable using a combination of features, e.g. by means of an LDA/SVM approach.

\begin{figure}[htbp]
\centering
\includegraphics[width=0.7\linewidth]{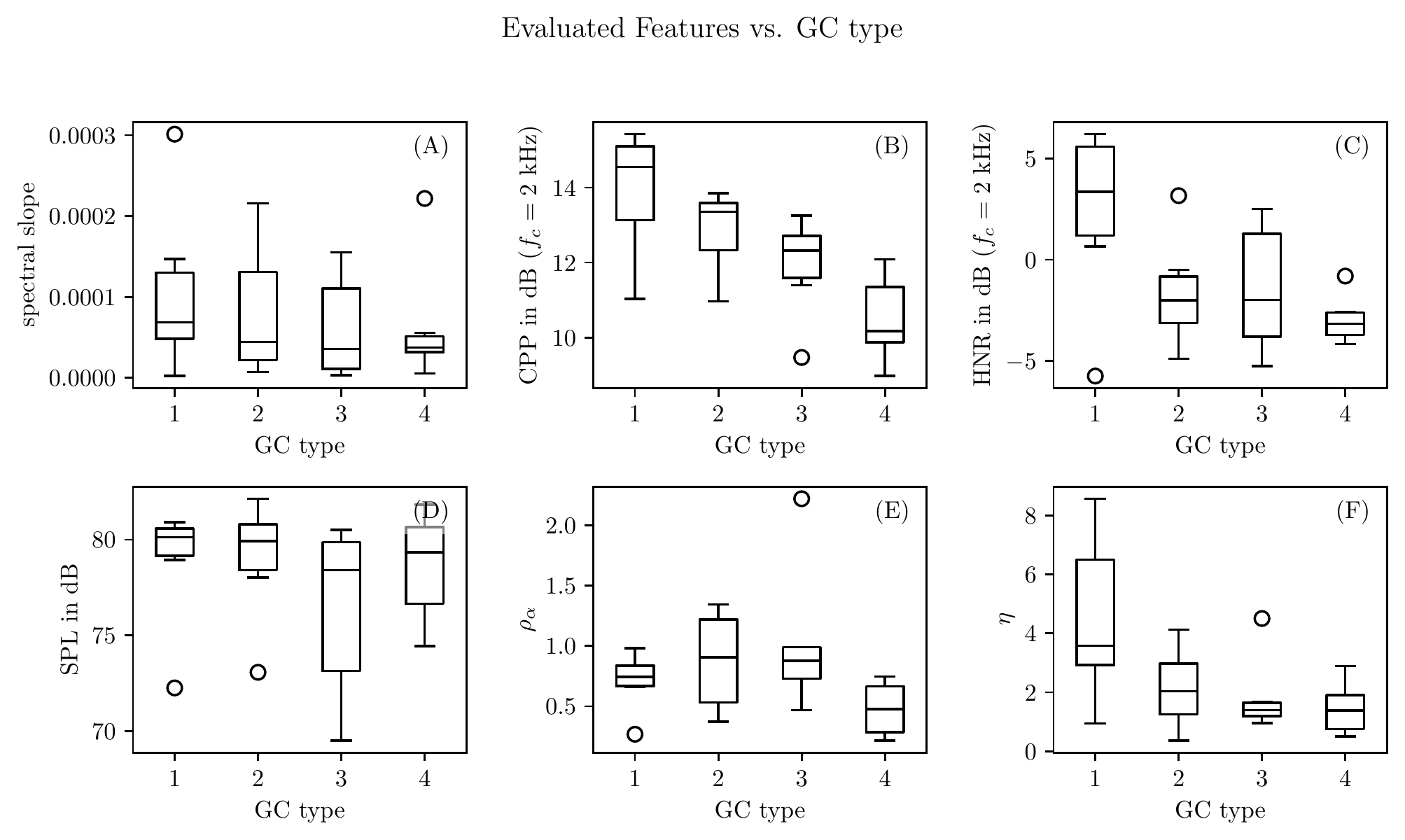}
\caption{Boxplot diagrams of the features spectral slope $a$, CPP $c_{\SI{2}{\kilo\hertz}}$, HNR $h_{\SI{2}{\kilo\hertz}}$, SPL $L_{P,\SI{5}{\kilo\hertz}}$, alpha ratio $\rho_\alpha$, and HBI $\eta$ as a function of glottal closure type for acoustic pressure signals at the microphone position.}
\label{fig:03-vis-acouPress-periodic-boxplot-GC}
\end{figure}

\paragraph{Boxplots for Glottal Closure Type}
In Fig.~\ref{fig:03-vis-acouPress-periodic-boxplot-GC}, the boxplot diagrams for the simulation parameter \textit{glottal closure type} are depicted for the acoustic pressure evaluated at the microphone position.
From Fig.~\ref{fig:03-vis-acouPress-periodic-boxplot-GC}~(A), (B), and (C), we see a general trend of decreasing CPP and HNR values with increasing initial glottal openings. A significant trend is visible for CPP evaluated with the low-pass filter at $f_c=\SI{2}{\kilo\hertz}$, as shown in Fig.~\ref{fig:03-vis-acouPress-periodic-boxplot-GC}~(B).

\begin{figure}[htbp]
\centering
\includegraphics[width=0.7\linewidth]{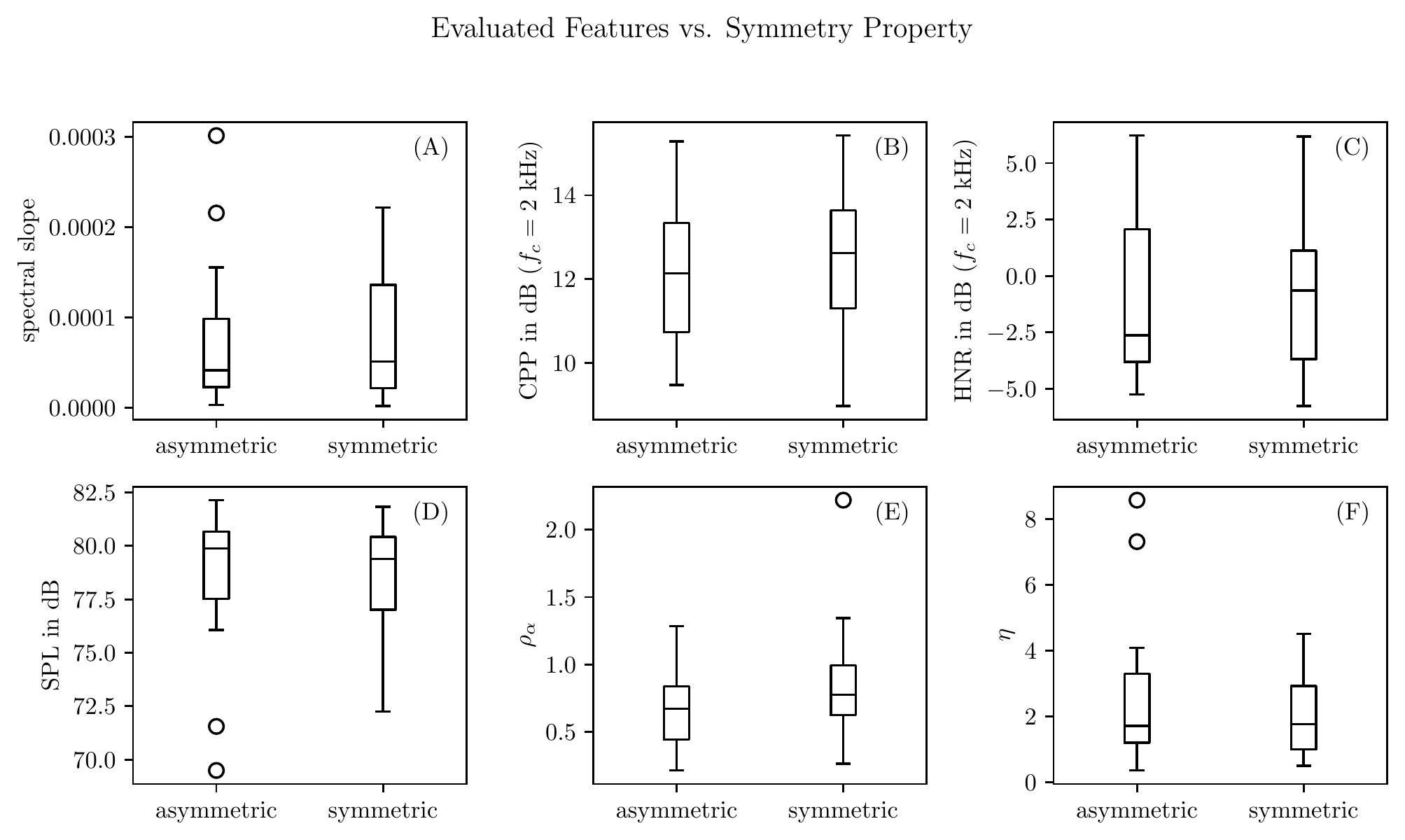}
\caption{Boxplot diagrams of the features spectral slope $a_{\SI{2}{\kilo\hertz}}$, CPP $c_{\SI{2}{\kilo\hertz}}$, HNR $h_{\SI{2}{\kilo\hertz}}$, SPL $L_{P,\SI{5}{\kilo\hertz}}$, alpha ratio $\rho_\alpha$, and HBI $\eta$ as a function of VF motion symmetry for acoustic pressure signals at the microphone position. }
\label{fig:03-vis-acouPress-periodic-boxplot-symmetry}
\end{figure}

\paragraph{Boxplots for Symmetry}
In Fig.~\ref{fig:03-vis-acouPress-periodic-boxplot-symmetry}, the boxplot diagrams for the simulation parameter \textit{symmetry} are depicted for the acoustic pressure evaluated at the microphone position.
It is shown, that no significant discrimination of glottal symmetry is possible based on the evaluated features. This is also supported by near-zero correlation values for the symmetry property depicted in Fig.~\ref{fig:03-vis-acouPress-periodic-corrMap}. This non-observable difference might be due to the inhomogeneous composition of the data.

\begin{figure}[htbp]
\centering
\includegraphics[width=0.7\linewidth]{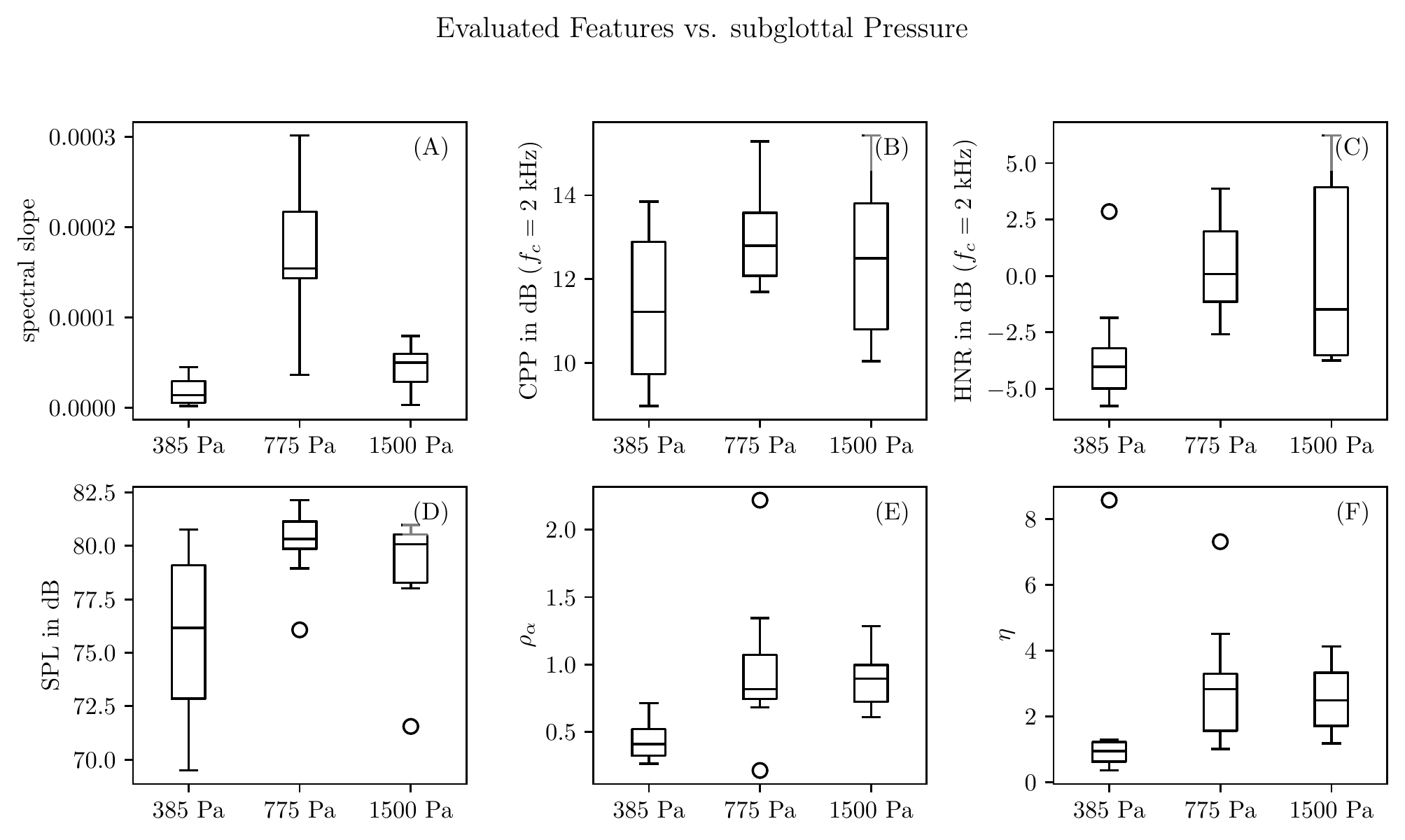}
\caption{Boxplot diagrams of the features spectral slope $a$, CPP $c_{\SI{2}{\kilo\hertz}}$, HNR $h_{\SI{2}{\kilo\hertz}}$, SPL $L_{P,\SI{5}{\kilo\hertz}}$, alpha ratio $\rho_\alpha$, and HBI $\eta$ as a function of subglottal pressure for acoustic pressure signals at the microphone position.}
\label{fig:03-vis-acouPress-periodic-boxplot-subglottPressure}
\end{figure}

\paragraph{Boxplots for Subglottal Pressure}
In Fig.~\ref{fig:03-vis-acouPress-periodic-boxplot-subglottPressure}, the boxplot diagrams for the simulation parameter \textit{subglottal pressure} are depicted for the acoustic pressure evaluated at the microphone position.
A discrimination between the subglottal pressures \SI{385}{\pascal} and $\{ \SI{775}{\pascal},\SI{1500}{\pascal} \}$ is possible using HNR, SPL, Alpha Ratio $\rho_\alpha$, and HBI $\eta$, as depicted in Fig.~\ref{fig:03-vis-acouPress-periodic-boxplot-subglottPressure}~(C), (D), (E), and (F), respectively. Spectral slope allows distinguishing configurations with \SI{775}{\pascal} from $\{ \SI{385}{\pascal},\SI{1500}{\pascal} \}$, as it can be seen from Fig.~\ref{fig:03-vis-acouPress-periodic-boxplot-subglottPressure}~(A).

\subsubsection{Intermediate conclusions boxplots}

From the boxplot diagrams, statements regarding the significance of the features' discrimination ability can be drawn. A good discriminative ability is provided by using CPP or HNR to discriminate the GC type. The decreasing trend of CPP for increasing glottal insufficiencies reported in Fig.~\ref{fig:03-vis-acouPress-periodic-boxplot-GC}~(B) is consistent with \cite{Hillenbrand1996, Birk2017, Thornton2019, Zhang2013Acoustic}. In accordance with \cite[Fig.~12]{Falk2021}, where a distinction of VF motion asymmetries was not evident based on their evaluated features, the acoustic features of the present work also do not allow for the discrimination of VF motion asymmetry. Similar to \cite{Zhang2013Acoustic, Falk2021}, we conclude that asymmetries in the VF motion are not necessarily related with a reduction in voice quality. The numerical range of our results is similar to the results of \cite{Falk2021}.
Furthermore, the subglottal pressure can be discriminated into two groups, i.e. \SI{385}{\pascal} and $\{ 775, 1500\}\SI{}{\pascal}$ by using the features SPL, alpha ratio and HBI, mainly.

\FloatBarrier

\subsection{LDA/SVM Classification}
\label{sec:03-results-LDA-SVM}


In the following figures, the background is colored according to the class boundaries evaluated with SVM. The coloring of the data points indicates if the SVM is able to classify the point correctly: the inner dot represents the true class, and the outer ring represents the predicted class by SVM.
From these figures, we see how the simulation configurations cluster in the dimension-reduced feature space. Furthermore, a graphical indication of class boundaries is provided. The class boundaries are obtained by classifying points of a mesh grid in the two-dimensional space with the trained SVM classifier.

\begin{figure}[htb]
\centering
\includegraphics[width=0.7\linewidth]{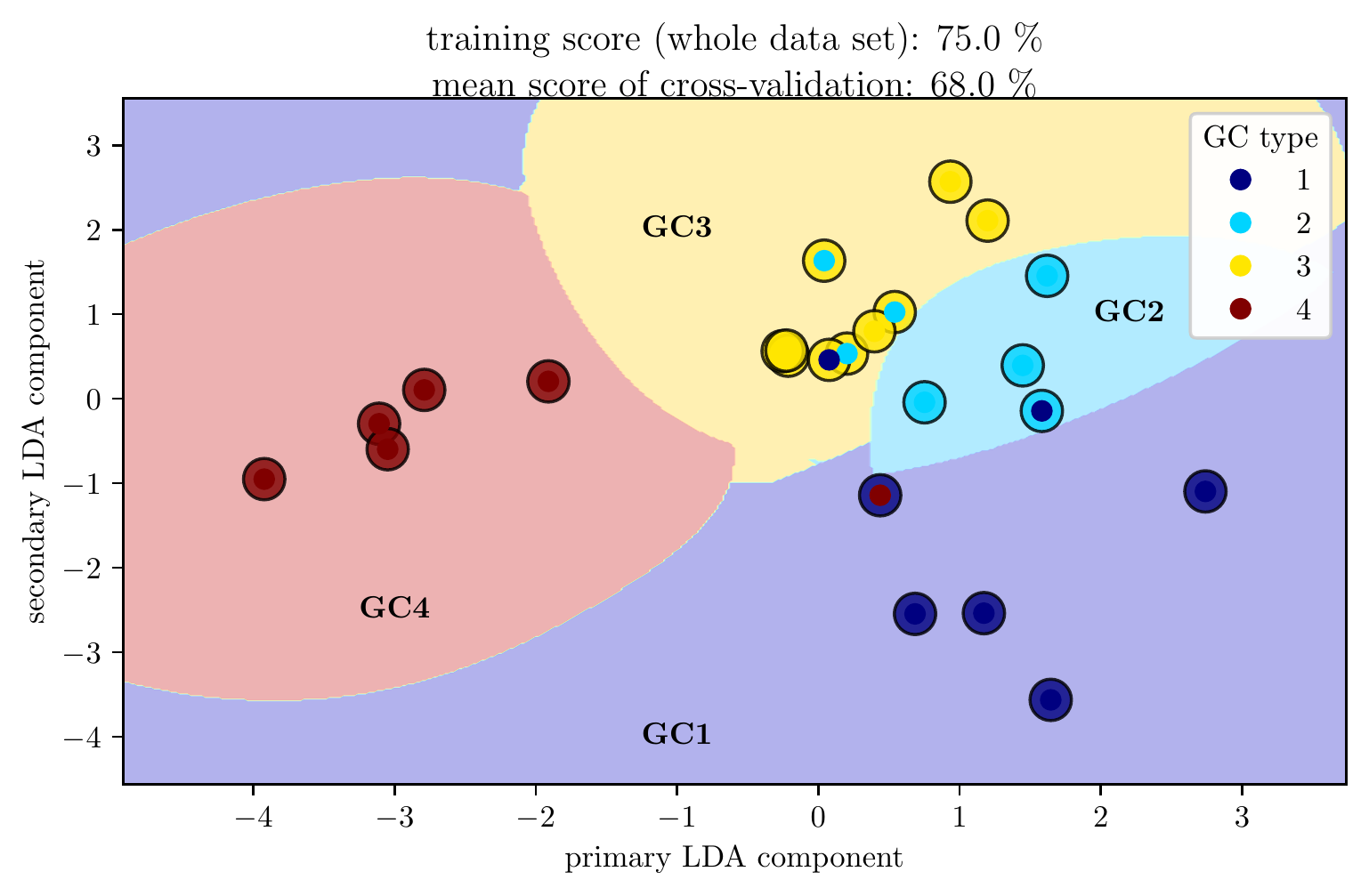}
\caption{(color online) LDA-Reduced dimensionality of feature space and 2D-SVM classification of the glottal closure type for acoustic pressure signals at the microphone position. The points are colored according to the true GC type. The colored ring around each point denotes the predicted GC type. The background colors show the predicted GC types of the dimension-reduced feature space as labeled, and thus the class boundaries.}
\label{fig:03-vis-acouPress-periodic-LDA-GC}
\end{figure}

\paragraph{LDA \& SVM for Glottal Closure Type}
In Fig.~\ref{fig:03-vis-acouPress-periodic-LDA-GC}, the result of the dimensionality reduction with LDA is visualized for all configurations using the pressure signal evaluated at the microphone point with respect to \textit{glottal closure type}. 

From Fig.~\ref{fig:03-vis-acouPress-periodic-LDA-GC}, we see that a clear separation is possible between GC4 and $\{ \mathrm{GC1}, \mathrm{GC2}, \mathrm{GC3} \}$. The classes GC1, GC2, and GC3 are interwoven in the 2D-space. Thus, the training score\footnote{The training score is obtained as follows. First the SVM is trained with all 24 configurations. Then the SVM is evaluated at the data points and the percentage of true classifications in relation to the total data is the training score. } is only \SI{75}{\percent}. Misclassifications occur primarily for GC2 and GC3.


\begin{figure}[htbp]
\centering
\includegraphics[width=0.7\linewidth]{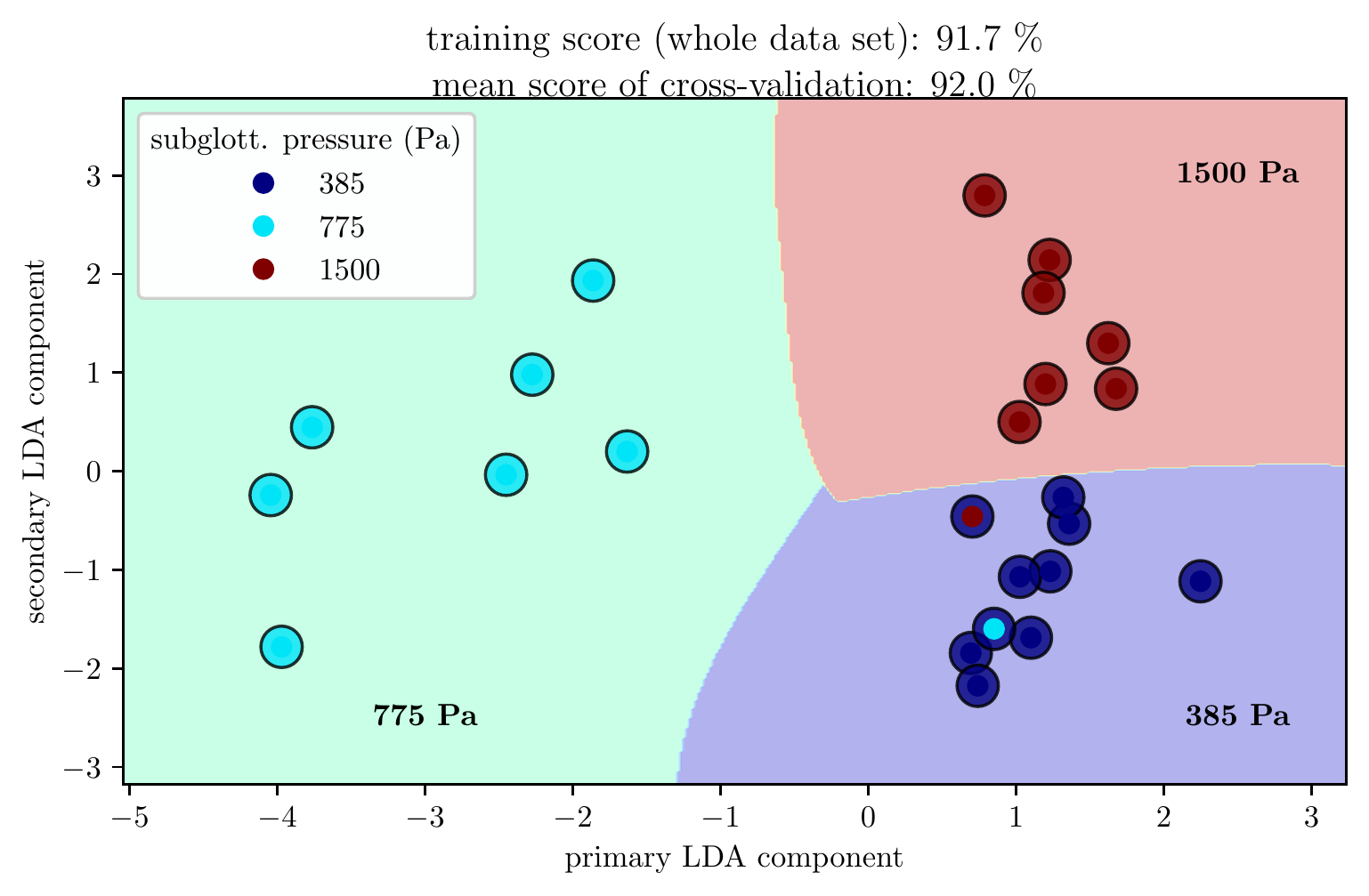}
\caption{LDA-Reduced dimensionality of feature space and 2D-SVM classification of the subglottal pressure for acoustic pressure signals at the microphone position. The points are colored according to the true subglottal pressure. The colored ring around each point denotes the predicted subglottal pressure. The background colors show the predicted subglottal pressure class of the dimension-reduced feature space as labeled, and thus the class boundaries.}
\label{fig:03-vis-acouPress-periodic-LDA-subglottPress}
\end{figure}

\paragraph{LDA \& SVM for Subglottal Pressure}
In Fig.~\ref{fig:03-vis-acouPress-periodic-LDA-subglottPress}, the result of the dimensionality reduction with LDA is visualized for all configurations using the pressure signal evaluated at the microphone point with respect to \textit{subglottal pressure}.
In Fig.~\ref{fig:03-vis-acouPress-periodic-LDA-subglottPress}, it is visible that a good class separability is given for the three subglottal pressures, which is resembled in the training score of \SI{91.7}{\percent}. Especially the configurations with \SI{775}{\pascal} are well separated from the others.

\begin{figure}[htbp]
\centering
\includegraphics[width=0.7\linewidth]{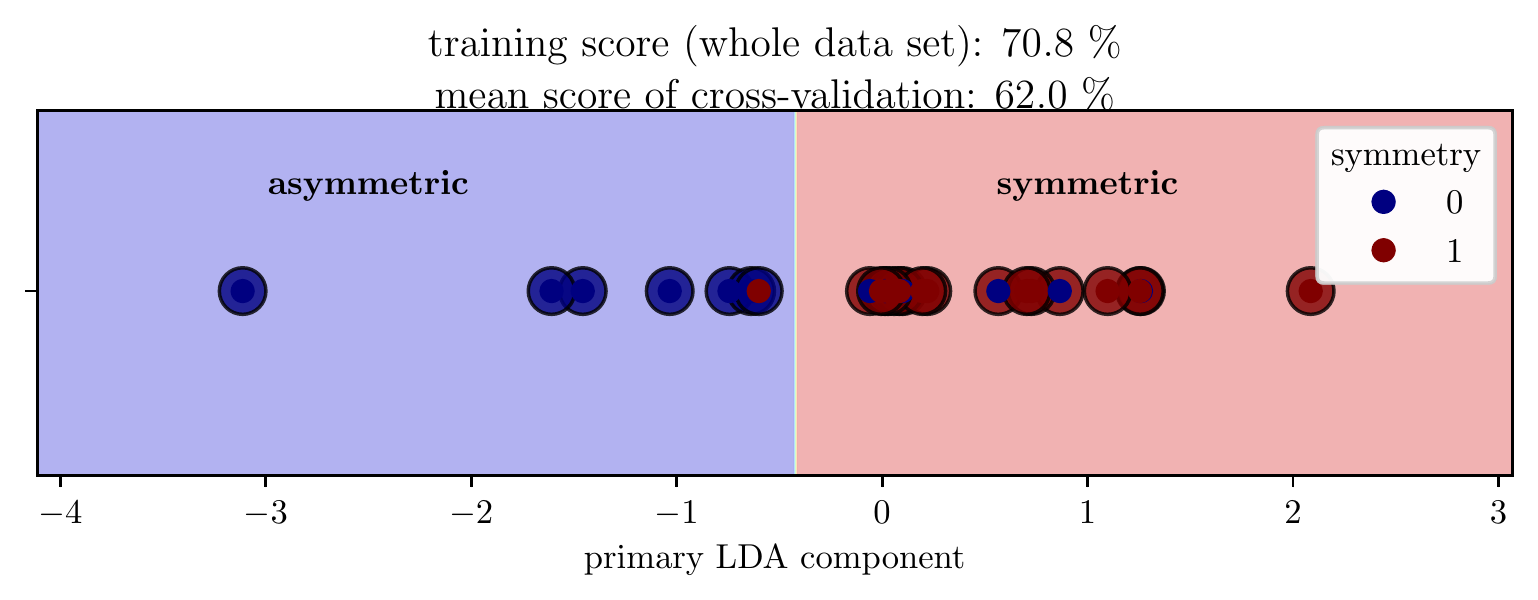}
\caption{LDA-Reduced dimensionality of feature space and 2D-SVM classification of the symmetry for acoustic pressure signals at the microphone position. The points are colored according to the true symmetry class. The colored ring around each point denotes the predicted symmetry class. The background colors show the predicted symmetry class of the dimension-reduced feature space as labeled, and thus the class boundary.}
\label{fig:03-vis-acouPress-periodic-LDA-symmetry}
\end{figure}

\paragraph{LDA \& SVM for Symmetry}
In Fig.~\ref{fig:03-vis-acouPress-periodic-LDA-symmetry}, the result of the dimensionality reduction with LDA is visualized for the pressure signal evaluated at the microphone point the microphone position with respect to the \textit{symmetry} property.
From Fig.~\ref{fig:03-vis-acouPress-periodic-LDA-symmetry} we might conclude that a good class separability is given because the class boundary is clearly between two clusters. Still, there are many data samples located at the wrong side of the class boundary which is obvious from the training score of \SI{70.8}{\percent}.

\FloatBarrier
\section{Conclusion}
\label{sec:04-conclusion}



Individuals affected by voice disorders have a significant reduction of their quality of life. Therefore, a simulation model \textit{simVoice} was developed to investigate typical characteristics of voice such as sub-glottal pressure and of functional voice disorders as glottal closure insufficiency and left-right asymmetry. Based on 24 different voice configurations, we study voice parameters (HNR, CPP, ...) in the prospect of analyzing the flow-induced noise generation. Although multiple studies are based on experimental data, this is one of the first studies learning simulated data of healthy and disordered voice signals. A major limitation of our investigations is that the dataset is small, but the definition of relevant characteristics are precise and based on the simVoice model.
Furthermore, the small dataset is analyzed by correlation analysis, and an SVM classifier with RBF kernel. With the use of LDA the dimensions of the individual representations are visualized in 2D. This allows to draw first correlations and determine the most important features evaluated from the acoustic signals in front of the mouth. Based on the definition of the acoustic features and with focus on the two lower formants, we found that a low-pass filtering with a cut-off frequency of \SI{2}{\kilo\hertz} provides a reasonable basis for further analysis.
Regarding the results, conclusions can be drawn as follows. The GC type can be best discriminated based on CPP. The SPL allows discriminating subglottal pressure into two groups. The glottal symmetry does not have any effect on the evaluated features.
Using the LDA-dimensionality-reduced feature space, one can best classify subglottal pressure with 91.7\% accuracy. In addition to that the GC type can be classified with 75\% accuracy. Finally, the symmetry property exhibits interwoven clusters and hence a classification is rather inaccurate. 



\section*{Acknowledgment}
We acknowledge support from the Austrian Academy of Sciences (ÖAW), research grant "Understanding voice disorders", received from "Dr. Anton Oelzelt-Newin'sche Stiftung".

\bibliography{references_BIBTEX}

\end{document}